\documentclass[superscriptaddress,prl,showpacs,preprint]{revtex4}
\usepackage{amsmath}
\usepackage{amsfonts}
\usepackage{graphics}
\begin{document}

\title{Jamming during the discharge of granular matter from a silo}

\author{Iker Zuriguel}
\email[Author to whom correspondence should be addressed. E-mail: ]{iker@fisica.unav.es}
\affiliation{Departamento de F\'{\i}sica y Matem\'atica Aplicada, Facultad de Ciencias,
Universidad de Navarra, E-31080 Pamplona, Spain.} 

\author{Angel Garcimart\'{\i}n} 
\affiliation{Departamento de F\'{\i}sica y Matem\'atica Aplicada, Facultad de Ciencias,
Universidad de Navarra, E-31080 Pamplona, Spain.} 

\author{Diego Maza}
\affiliation{Departamento de F\'{\i}sica y Matem\'atica Aplicada, Facultad de Ciencias,
Universidad de Navarra, E-31080 Pamplona, Spain.} 

\author{Luis A. \surname{Pugnaloni} \footnote{Current address: Instituto de F\'{\i}sica de L\'{\i}quidos y Sistemas
Biol\'ogicos, CONICET-UNLP, cc 565, (1900) La Plata, Argentina.}}
\affiliation{Procter Department of Food Science, 
University of Leeds, Leeds LS2 9JT, United Kingdom.}

\author{J.M. Pastor} 
\affiliation{Departamento de F\'{\i}sica y Matem\'atica Aplicada, Facultad de Ciencias,
Universidad de Navarra, E-31080 Pamplona, Spain.}

%\date{}

\vspace{1cm}
\begin{abstract}

\noindent

In this work we present an experimental study of the jamming that stops the free flow of grains from
a silo discharging by gravity. When the outlet size is not much bigger than the beads, granular material 
jams the outlet of the container due to the formation of an arch. Statistical data from the number of grains fallen between consecutive jams are presented. The information that they provide can help to understand the jamming phenomenon. As the ratio between the size of the orifice and the size of the beads is increased, the probability that an arch blocks the outlet decreases. We show here that there is a power law divergence of the mean avalanche size for a finite critical radius. Beyond this critical radius no jamming can occur and the flow is never stopped.
The dependence of the arch formation on the shape and the material of the grains has been explored. It has been found that the material properties of the grains do not affect the arch formation probability. On the contrary, the shape of the grains deeply influences it. A simple model to interpret the results is also discussed.

\pacs{45.70.Ht, 45.70.Mg}
\end{abstract}

\maketitle

\newpage
%%%%%%%%%%%%

\section{Introduction}

We often experience that things get jammed. We can easily recall many situations where a flow of discrete particles is arrested, and this we call -- in a broad sense -- a jam. Recently this concept has been generalized by A. Liu and S. Nagel \cite{Liu} introducing the idea of ``jamming transition'', characterized by the sudden arrest of the particle dynamics. Glass transitions, colloidal gels, foams, as well as granular flows, traffic and stampedes are included among the systems that display jamming.

``Fragile matter'' is another way to describe these systems \cite{cates}. Qualitatively, we are concerned with fragile matter when a system is mechanically stable regarding the so-called compatible stresses, and unstable against uncompatible stresses. Maybe it is more intuitive to consider that jamming is the response of a system to the applied external stresses by developing mechanical structures that block the flow. 

In granular materials, these structures are called ``arches''. Arching is one of the most important characteristics of granular materials. Arches are responsible for the non-uniform stress propagation \cite{fuerzas} and for changes in the volume fraction \cite{nowak} because they create voids. In practice, for monodispersed spheres, the maximum reachable value for the volume fraction in $3D$ is $0.64$, corresponding to the \emph{``random close packing''}, although the definition of this state is not clear \cite{torquato}. Under the effect of the gravity the minimum value is $0.52$, corresponding to the \emph{``random loose packing''}. Arches have been studied experimentally \cite{to}, reproduced with numerical simulations \cite{pugnaloni, pugnaloni2, manna} and analyzed with theories \cite{drescher}. An arch is an intrincate structure where the particles are mutually stable. If one of the particles that form an arch is removed, the arch collapses under the effect of gravity. A basic understanding of the physical mechanisms underlying arch formation is lacking. Hence the interest of adding an experimental investigation to the existing knowledge. Moreover, observation of arches in the bulk of the granular material is hardly accessible experimentally, which makes difficult the validation of theoretical models \cite{pugnaloni, pugnaloni2}. For this reason we need to study arch formation at the outlet of a silo as a first attempt to elucidate arch properties and the related jamming phenomenon in 3D.

In a previous work \cite{Nuestro} the arching at the outlet of a silo discharged by gravity was studied. The experience was simple: granular material flowing throw the outlet of a silo can jam if the size of the orifice is not big enough. An arch or a dome is formed and to restart the flow an input of energy (blowing, shaking or tapping) is necessary to break it. After that, the flow is restored until a new arch blocks the outlet. The size distribution of avalanches for a fixed $R$, where $R$ is the ratio between the orifice radius and the grain radius, was characterized. In order to explain such a distribution a simple model was proposed based on the assumption that the probability $p$ that a particle passes through the outlet without blocking it is independent of its neighbours. This model explained quite well the results obtained, but some aspects, such as the probability of having very small avalanches or the notion that arching is a collective event, were not included. It was also shown that $R$ is the only relevant parameter concerning size, in the sense that the absolute measure of the grain or the orifice is irrelevant provided that $R$ is the same.

%%%%%%
In this work we extend that study by using different grain materials and grain shapes in order to explore their influence on jamming. Surface roughness, restitution coefficient, density and other parameters are varied by using different materials. The influence of these properties in arch formation can provide interesting clues about the variables that do affect the phenomenon. This can help to understand whether arch formation is due to processes involving the physical properties of the material or only to their geometry. Besides, a wide range of $R$ is studied to determine whether a critical radius -- above which jamming is not possible -- does exist. 

This manuscript is structured as follows. First, we provide a detailed description of the experimental setup. We then present the data obtained in a typical run. The existence of a critical radius is discussed, as well as the influence of grain material and geometric properties. We introduce a modification of the model previously used that provides a better understanding of the results. Finally, some conclusions are drawn.
%%%%%%

\section{Experimental set-up}

The experimental set-up consists of a scaled cylindrical silo with a circular hole in the base (Fig.\ref{sketch}). When the silo is filled with granular material, grains pour freely from the outlet due to gravity. However, when the diameter of the orifice does not exceed a few bead diameters, the flow is soon arrested because of the formation of an arch. We weight the fallen mass of grains with an electronic balance placed beneath the silo. Knowing the weight of one grain, we determine the number of grains fallen between two successive jamming events. We call this event an avalanche. This measurement is stored in a computer. Then, the arch formed at the outlet is destroyed by means of a jet of pressurized air from beneath the orifice, and another avalanche is triggered.  

Silos of different diameters ($30$, $80$, $120$ and $150\, mm$) were used in order to explore the influence of this dimension on the jamming probability. We checked that if the silo diameter is larger than approximately $30$ bead diameters, the finite size of the bin can be neglected \cite{hirshfeld}. We have used silos made of two different materials (stainless steel and glass) to verify that the properties of the walls do not influence the results. All the silos are of the same height, namely, $500\, mm$. It is important to remark that the geometry of the container affects significantly the jamming. In their paper, K. To et al. \cite{to} explain that the jamming probability in a 2D silo remains constant for angles smaller than a critical value. We have used flat bottomed silos in order to be under this critical value, regardless of the kind of particle.  

The flat bottom of the silo is a disk with a circular hole in the center. More than 50 disks with different hole radius ($\phi$) have been used; disks of brass, steel and glass do not yield any noticeable difference in the results. The orifice is a nozzle that opens downward (see inset of Fig.\ref{sketch}). This is done in order to avoid jams at the very orifice. The hole radius is measured with a precision of $0.05\, mm$. 

We have used several types of granular materials (see Table 1). Beads with different radius, $r$, were used to check that the dimensionless radius $R$ ($R = \phi/r$) is indeed the control parameter. Grains of different materials (sets 2, 6, 8 and 9) were used in order to explore the influence of their properties on the jamming. We have also carried out several runs with glass beads with high size dispersion (set 4 in table 1). The effect of surface roughness was also investigated. We changed the roughness of the glass beads surface with a chemical treatment with fluorhidric acid. We repeated this treatment several times with different concentrations of fluorhidric acid ($1\%, 2\%, 5\%, 10\%, 20\%$ in concentration) and a qualitative change of the surface roughness was observed with both a magnifying glass and an electronic microscope. The properties of the particles after the treatment with fluorhidric acid ($20\%$ in concentration) are presented in the set 5 of the Table 1 and a photograph is shown in figure \ref{sis_exp2}b. In addition, the effectiveness of the roughing method was asessed by observing the beads roll on a smooth surface: initially they followed straight paths; after the treatment, the paths were irregular. Note that while the surface roughness of the particles was changed, the granular friction coefficient remained approximately the same. The granular friction coefficient can be determined by measuring $\theta$, the angle of a pile when a surface avalanche is developed. The higher $\theta$, the higher the intergranular friction coefficient is \cite{Svarovsky}. It is shown in Table 1 that this angle is not considerably affected by the surface roughness of the particles. Finally, we have used grains with shapes departing from the sphere: pasta grains -- see a sketch of their shape in Fig. \ref{sis_exp2}a -- lentils and rice (sets 9, 10 and 11 respectively). Most of the results were obtained with glass beads with low dispersion in size (sets 1 and 2 in the table). From now on we will refer to these glass beads unless otherwise is specified.

When the grains fall through the outlet they are collected in a cardboard box on a balance (Sartorius GP 4102), which has a resolution of $0.01\, g$. This resolution allows us to detect a single bead in all cases (see Table 1) except for the smallest glass beads, steel and pasta grains (sets 3, 8 and 9 respectively in the table). For these last three sets the smallest detectable amount is $8$, $3$ and $2$ grains respectively. 

Since the silo filling procedure is known to have a marked effect on the flow pattern developed during the discharge \cite{zhong}, we have tried to fill the silo always in the same way. We have used the ``concentric filling procedure'', which is accomplished by rapidly pouring the grains at the center of the silo. With this method, a mixed flow is developped when spheres are used. In this flow, every sphere remains in the same position relative to its neighbours in the upper part of the silo. Below a certain point, the spheres near the wall move slower than the spheres in the bulk \cite{zhong}. In mixed flows, there can be a stagnation region near the bottom corner; but in our experiment we have observed that even this beads move toward the orifice. We have seen it using transparent glass silos. We have observed that when the granular material inside the silo decreases to a level of about 1.2 times the diameter of the silo, a funnel flow develops. In order to avoid this, the silo is refilled from time to time.  In runs with lentils or rice, the flow pattern developed inside the silo is always a funnel flow (the grains near the wall move slower than the grains in the center in the whole silo).

Due to the Janssen effect \cite{duran}, the pressure at the bottom of the silo remains constant during the experiment provided that the level of the material exceeds roughly 1.5 times the diameter of the silo. The volume fraction of the material has been measured in different situations. For spherical beads and pasta grains we have obtained a volume fraction of about $0.59 \pm 0.02$, and this value remains almost constant during the discharge. For lentils the volume fraction is $0.59 \pm 0.02$ and it is $0.55 \pm 0.02$ for rice, but in these last two cases it is not possible to check the volume fraction during the discharge due to the funnel flow.

%Since the silo filling process has been seen to have a marked effect on the flow pattern developed during the discharge \cite{zhong}, we have tried to fill the silo always with a concentric filling procedure to obtain a mixed flow. In this flow, every grain remains in the same position respect to its neighbours in the upper part of the silo. However, the grains near the wall below the ``effective transition point'' -- the point at which the internal flow channel boundary reaches the wall \cite{zhong} -- move slower than the grains in the bulk. Using transparent glass silos we have seen that all the grains (even the grains in the bottom corner) move toward the orifice. The concentric filling procedure is accomplished by rapidly pouring the grains at the center of the silo. The compaction of the material has been measured in different situations. For spherical beads and pasta grains we have obtained a volume fraction of about $0.59 \pm 0.02$. This value remains almost constant during the discharge. In the case of lentils the volume fraction is $0.59 \pm 0.02$ and $0.55 \pm 0.02$ in the case of rice. In these last two cases is not possible to check the volume fraction during the discharge because the flow pattern developed during the discharge inside the silo is purely a funnel flow (in the whole silo the grains near the wall move slowly that the grains in the bulk).

After an arch is formed at the outlet, we resume the flow by means of a jet of pressurized air aimed at the orifice. This mechanism has been chosen to avoid possible changes in the volume fraction of the granular material \cite{nowak} created by other mechanisms generally used in the industry (vibration of the silo or hitting the wall). The blow is controlled by shortly opening an electrovalve. This valve is driven by a switch that is in turn controlled from a PC. We have observed that as long as the arch is destroyed, the time during which the air is blowing and the air pressure do not affect significantly the results. Usually, we kept the air pressure at $4 \pm 0.5$ atmospheres and the air jet lasts $0.4 \pm 0.1$ seconds. 

%an SMC SY5120-6LOU-01F-Q electrovalve
%in a Hewlett-Packard HP34970A multimeter

%The elapsed time of each avalanche is measured with a microphone placed next to the hole that picks  the noise of the falling beads. The signal is amplified and sent to a digitizing oscilloscope (HP54510A), and then analyzed to obtain the duration of the avalanche. The resolution is of about 0.1 seconds. As we know the fallen mass and the duration of each avalanche, the flow can be computed. Work is under way and the results about the flow will be published elsewhere.

%with a Vlapey-Fisher VP-1093 Pinducer microphone

We have monitored the room temperature and the humidity: the room temperature was kept almost constant, at about $22\pm2$ $^oC$ and the relative humidity fluctuates between 35 and 60 \%. Due to the heavy weight of the particles, cohesive and electrostatic forces can be neglected when compared to gravity.

%with a Rotronic Hygropalm 2

%All the instruments are connected to the computer through a GPIB and a RS-232 bus, and the measurement process is automated. The flow diagram of the computer program is as follows: a blow of pressurized air is released by opening the valve in order to trigger an avalanche; when the avalanche ends, the weight of the grains is measured and converted into number of grains. The measurements are time stamped and stored, and after a short waiting time the procedure is repeated. When the level of the grains decreases below the preset height of two silo diameters, the program stops; then the silo can be refilled and the program resumed. Typically, several thousand avalanches are registered in each run.

\section{Experimental results}

\subsection{Avalanche size distribution}

The avalanche size distribution $n_R(s)$ -- the normalized histogram -- for a fixed value of $R$ is represented in Fig.\ref{jam_4histos}a. We define the size of the avalanche, $s$, as the number of grains fallen between the
air blow and the jamming of the outlet. The typical number of avalanches that we collect to make an histogram for a given $R$ is about $3000$. Sometimes, when more resolution is needed, up to $100000$ avalanches are registered.

%The procedure followed to construct a histogram is the following. For a fixed $R$, if the biggest avalanche obtained is smaller than around $100$ beads, we select the bins so that each one corresponds to one bead. If the biggest avalanche is greater than $100$ beads, we make the histogram with a number of bins between $50$ and $100$. Only one condition is imposed to make the histogram: each bin must correspond to an integer number of particles. One of the problems we have found to make an histogram is related to the resolution of the balance. Usually the number of beads obtained by dividing the fallen mass by the bead weight is non-integer. When rounding these values to construct the histogram, discrete jumps appear which are smoothed using dithering \cite{ruido}. This problem is only important for small $R$, when the size of the bin corresponds to just one or two beads.

In the histogram, two different regions are observed (Fig.\ref{jam_4histos}a). Typically, the number of avalanches smaller than the mode (the avalanche size with the highest probability) grows with $s$. It seems that the dependence follows a power law (Fig.\ref{jam_4histos2}a). However, the region of small avalanches is too small to assert conclusively this result. The number of big avalanches (larger than the mode) decreases exponentially (Fig.\ref{jam_4histos2}b). This behaviour is found in almost all histograms for all the values of $R$. 

For avalanches larger than the mode, the probability that a grain gets jammed is constant (it remains the same at all times). The exponential tail of the histogram is an evidence that arch formation is an uncorrelated process, and that $p$ --  the probability that one grain gets past through the outlet without blocking it -- is unconditional in the sense that it is constant during the course of an avalanche. Furthermore, the first return map shows no sign of correlation between consecutive avalanches (Fig.\ref{jam_4histos}b).

The first part of the histogram (small size avalanches) is not well understood. In fact, this region is very sensitive to parameters like the duration of the air jet, the air pressure and the diameter of the silo. For example, we have carried out experiments with air pressure spanning from 1.5 to 12 atmospheres and it is observed that this variable has an influence on the shape of the histogram in this region. 

We have found that the features of the growing region of $n_R(s)$ depend strongly on $R$. In the limit of small $R$ ($R<1.6$) the mode corresponds to $s=1$ and consequently the growth zone does not exist. For large $R$, this region is difficult to study because the bins of the histogram are larger and the growing portion is hidden in the few first points. Many avalanches are required to obtain a good resolution in the histogram (around $30000$ avalanches were needed for Fig.\ref{jam_4histos2}a). We have obtained the mode (the value of $s$ where the histogram peaks) whenever possible; the result is shown in the inset of Fig.\ref{jam_4histos}a. As explained, the error bars grow larger and larger as $R$ increases, until they eventually are so important that they render the measurement meaningless. We cannot therefore describe the behaviour of the mode as a function of $R$ for a wide range of $R$. 

Using a transparent glass disk at the bottom of the silo, visual inspection suggests that the cause of the growth of $n_R(s)$ with $s$ for small $s$ is the following. When a bridge is broken there is a minimum number of beads that fall through the hole. There is a high probability that, when the bridge is broken, all the particles that were forming it will fall without blocking the outlet. That is, the initial perturbation creates a transient flow whose features differ from the steady state reached later on during the course of an avalanche. If this is the case, the mode would grow with a power of $R$; indeed the data of the inset of Fig.\ref{jam_4histos}a can be fitted with a power of $R$ (the exponent is between 3 and 4), but the error bars are so large that this result is not conclusive.

In short, there exist two regions in the avalanche size distribution: the small increasing part of the $n_R(s)$ for avalanches smaller than the mode, associated to the occurrence of a transient regime (very sensitive on $R$ and on the unjamming procedure) and the exponential tail that corresponds to a steady flow where the jamming probability is time-independent.

\subsection{The existence of a critical radius}

The exponential tail of the histogram for spherical beads is always present regardless of $R$. The characteristic exponent depends strongly on $R$: the bigger $R$, the smaller the exponent is. In our experiments, the range in the size of avalanches is very wide. For $R=1.32$ the avalanche size distribution goes from $1$ to $10$ particles whereas for $R=4.31$ it goes from $100$ to $600000$. As the histogram is mostly exponential, we introduce a characteristic parameter in the exponent to rescale all the histograms in order to compare them. We have rescaled the avalanche size $s$ with the mean avalanche size $<s>$ corresponding to each $R$. When the rescaled histograms are plotted, all of them collapse into a single curve (Fig.\ref{histos_nor}). Clearly, the collapse is not perfect due to the effect of the first part of the histogram, which is not exponential. However, at least for the exponential tail, the histogram can be characterized with only one parameter, namely, the mean avalanche size $<s>$. 

One interesting question is whether there exists a critical radius $R_c$ above which jamming is not possible. Several papers in the field of engineering have been published about this topic, but the critical radius is approached from values of $R$ larger than $R_c$ \cite{Brown, Bever}. These previous works suggest a value for the critical radius roughly between $R=5$ and $R=10$. Other recent works have been done, but it is not yet clear if a critical radius actually exists and, if so, what its value is \cite{Nuestro, to}. 

In order to investigate this point, we have obtained the histogram for about 50 different values of $R$. As there is just one parameter which completely characterizes the histogram, $<s>$, it is reasonable to investigate the behaviour of this variable as $R$ is changed. The mean avalanche size $<s>$ is plotted versus $R$ in Fig.\ref{s_vs_r}. It is found that the data are well fitted by a power law:

\begin{equation}
<s>=\frac{A}{(R_c-R)^{\gamma}}
\label{eq3}
\end{equation}
%%%%

where $\gamma=6.9 \pm 0.2$, $A$ (which is a constant that corresponds to the value of $<s>$ when $R_c-R = 1$) is $9900 \pm 100$ and $R_c = 4.94 \pm 0.03$. In the inset of Fig.\ref{s_vs_r} the same results are plotted versus $1/(R_c-R)$ in a logarithmic scale. The power law divergence is clearly observed in this figure. It is worth noticing that the power law fit holds even far from $R_c$. The critical radius obtained for spherical beads is in the lower limit of the interval suggested in the literature. The unusually high value of the exponent $\gamma$ is somewhat surprising.

%This is not usual in phase transitions, although there are some cases -- like in Rayleigh-B\'enard convection -- where the same has been observed \cite{Rayleigh-Benard}.

\subsection{The effect of grain properties}

As mentioned above, the comparison of the results with changing material properties and grain shape can provide interesting clues to the jamming process. Among the material properties, those affecting density, elasticity and the surface roughness are the most immediately attractive to explore. Apart from glass, we have carried out several runs with beads of different materials (see Table 1). The surprising result is that all these changes do not produce any measurable effect in $<s>$ (Fig.\ref{diver_mat}a). This means that jamming is not directly related to the details of density or the elasticity of the material, nor to the surface properties of the particles. This last results seems to be in contradiction with the result reported by To et al.; they mention that disks with smooth edges jam with different probability than disks with rough edges \cite{to}. However, the rough edges in that experiment were formed by grooves of $0.2\; mm$ in depth. Such relatively large alteration of the surface may be
considered a change in shape rather than in the surface properties. In fact, To et
al. have seen ``concave'' arches promoted by disks that lock to each other by the
grooves.

A moderate size dispersion of the beads (12\% in radius) has also revealed itself to cause no effect whatsoever in the statistics. It is known that beads with a small size dispersion prevent the formation of of crystalline domains (clusters of locally ordered beads). Our results imply that local order of the grains was either already absent with monodisperse beads (the most probable assumption, as will be explained shortly) or that it does not influence the jamming phenomenon.

Let us now describe the effect of changing the shape of the particles -- departing from the spherical form -- on the jamming. To this end, we have carried out several runs with rice, and pasta grains (see Table 1). The results are compared with spherical glass beads in Fig.\ref{diver_mat}b, where the behaviour of $<s>$ as a function of $R$ is shown. It can be seen that the shape of the grains dramatically affects the jamming phenomenon. It is noteworthy that there is a critical radius even for grains that are not spherical. Its value changes depending on the shape ($R_c=5.03 \pm 0.05$ for pasta grains and $R_c=6.15 \pm 0.08$ for rice), but the divergence can always be fitted with Eq. 1. The inset of Fig.\ref{diver_mat}b shows that the critical exponent $\gamma$ remains approximately the same. From the results obtained with spherical beads, pasta and rice we can establish, qualitatively, that the nearer to the spherical shape, the bigger $<s>$ is and the lower $R_c$ is. 

The study of the form of the histogram for different granular materials is very interesting. The data show that there is no significant difference between the histograms for different materials, provided that the shape of the grain is spherical (Fig.\ref{histos_mat}a). Small differences can only be appreciated in the first part of the histogram, i.e. for avalanches smaller than the mode. These results show that, for avalanches bigger than the mode, $n_R(s)$ is independent of the material density and texture. 

A small change in the shape of the grains cause a great change in the form of the histogram (Fig.\ref{histos_mat}b). In the cases of rice and lentils the exponential decay found for spherical grains is not obvious. Moreover, we have found that the form of the histogram depends strongly on $R$. The influence of the grain shape in the jamming merits further investigation but with these results we can already assert that it affects jamming more deeply than the properties of the material.

\section{A model for spherical grains}

In a previous paper \cite{Nuestro}, we introduced a simple model that can describe the exponential decay of $n_R(s)$ for large $s$. After a closer look at the data, we propose here a refinement of the model wich includes a first attempt to explain the behaviour of $n_R(s)$ at small $s$.

The behaviour of $n_R(s)$ for large $s$, let us call it $n'_R(s)$, can be modelled as follows. During the steady flow of a single avalanche, grains pass through the opening of the silo either sequentially (if the opening is rather small) or in groups (if the outlet is at least two-particle-diameter wide). In any case, in order to block the opening, a grain has to cooperate with other grains to form an arch. Arches can be found everywhere in the interior of a granular packing \cite{pugnaloni}. However, during the discharge of a silo, only those arches formed at the opening will arrest the flow of the entire granular sample. Let us consider a grain that is moving through the region just above the opening (the jamming zone) where formation of an arch can lead to total jamming. Let us assume that $(1-p(R))$ is the probability that this grain comes together with other grains to form a blocking arch. This probability decays with increasing radius $R$ of the opening since the formation of a blocking arch depends on the minimum lateral span needed to reach opposite edges of the opening (see for example Ref. \cite{to} for arches in two dimensions). Then, the probability for a grain to flow through the jamming zone and get pass the opening without forming part of a blocking arch is $p(R)$. Wherever we use $p$ in the rest of the paper the dependence on $R$ is implied.

If the formation of different structures at the opening is uncorrelated, then the probability $p(R)$ must be the same for every particle that moves through the jamming zone. Therefore, the probability of having $s$ grains that indeed fall down followed by a grain that becomes part of a blocking arch is $p^s(1-p)$. This expression corresponds to the probability for having an avalanche of $s$ grains, \textit{i.e.} 

\begin{equation}
n'_R(s) = p^s(1-p).
\label{e4} 
\end{equation}

A semi-logarithmic plot of $n'_R(s)$ will present a straight line with slope $log(p)$. This is precisely the type of behaviour that we observe in our experiments for large values of $s$ (see Fig.\ref{jam_4histos2}b). Then, the probability $p$ for a given value of $R$ can be obtained from the slope of the exponential tail of the experimental $n_R(s)$ in a log-linear plot. Following this procedure we obtained Fig.\ref{figurap}, where we plot $p$ as a function of $R$. A fit for this dependency is difficult to decide. However it is important to note that this figure is another form of presenting the experimental data shown in the figure \ref{s_vs_r}, because $p$ and $<s>$ are related (both are parameters that characterize the histogram).  

Let us focus now on the behaviour of $n_R(s)$ for small $s$. The experimental $n_R(s)$ shows an initial growth from $s=0$ up to a maximum at intermediate values of $s$. The position of this maximum shifts to higher values of $s$ as the relative size of the opening $R$ is increased. We explain this effect by realizing that the initial perturbation aimed to trigger an avalanche (the compressed air blow) originates a transient flow. Only after the beads have settled down, the steady state flow of the avalanche starts. Of course, the larger the opening, the larger the initial perturbation is, and the longer it takes the system to enter the steady state flow.

The number of grains $s$ that falls through the opening until the steady flow is reached follows a certain distribution $n''_R(s)$. We expect this distribution to be narrow and with a mean at small values of $s$ if $R$ is small ($R < 2$). However, $n''_R(s)$ should be rather broad and with a large mean value if $R$ is close to the critical non-jamming transition value $R_c$. In order to obtain an analytical expression later on -- when we combine the transient regime and the steady state regime of the avalanche -- we model $n''_R(s)$ using a simple exponential distribution, \textit{i.e.}

\begin{equation}
n''_R(s) = (1-e^{-1/\alpha})e^{-s/\alpha}.
\label{e5} 
\end{equation}

Here, $\alpha(R)$ is a positive number that depends on $R$. The mean of $n''_R(s)$ is $1/(e^{1/\alpha(R)}-1)$, which gives a measure of the number of beads that needs to flow through the opening before the steady state is achieved for an opening $R$.

The total avalanche distribution, $n_R(s)$, is given by the convolution of $n''_R(s)$ with $n'_R(s)$, since they describe two consecutive processes:

\begin{equation}
n_R(s) = \sum\limits_{k=0}^{s} n''_R(k) n'_R(s-k).
\label{e6} 
\end{equation}

Form Eqs. \ref{e4}, \ref{e5} and \ref{e6} we obtain

\begin{equation}
n_R(s) = \frac{(1-\beta)(p-1)}{\beta-p}(p^{s+1}-\beta^{s+1}),
\label{e7} 
\end{equation}

where $\beta(R)=exp(-1/\alpha(R))$. Notice that $\beta$ can take values only between 0 and 1 since $\alpha$ is positive. From the fitting of expression \ref{e7} to our experimental data (see below), we have seen that $p>\beta$. This implies that $n_R(s) \propto p^s$ for large $s$, which agrees with the behaviour of $n'_R(s)$.

The comparison between the experimental avalanche size distribution and Eq. \ref{e7} is shown in Fig.\ref{figura_luis}. We have fitted the value of $p$ so that the correct slope at large $s$ is obtained in a semi-log plot. Then, the value of $\beta$ is fitted to obtain the correct position of the mode. As we can see, the model is able to describe the avalanche distribution at least qualitatively. However, the exponential distribution is probably not the best model for the initial transient flow. As we mentioned, this initial process is very dependent on the duration of the compressed air blow and the air pressure; therefore, it might prove very difficult to model. However, it is clear that the existence of such transient flow can explain the grow of $n_R(s)$ with $s$ for small values of $s$.

\section{Conclusions}

In this paper we have looked into the jamming during the discharge of a silo through small orifices. We have first shown that for spherical beads the avalanche size distribution, $n_R(s)$ is well defined, with an exponential tail for avalanches bigger than the mode. This behaviour can be understood if each grain passes through the outlet with a mean probability $p$ that is constant during all the discharge. For avalanches below the mode we obtain a behaviour that is not yet well understood. It is sensitive to changes in the particularities of the experimental conditions. We propose that this behaviour could be originated by the existence of a transient flow. Moreover, we have presented a simple model that incorporates the main features of the phenomenon.

The form of the histogram corresponds to an exponential decay, except for the small values of $s$, provided that the beads are spherical. When the shape of the grains is not spherical, the form of the histogram changes considerably. In this case, the shape of the histogram depends strongly on $R$. 

We have shown the existence of a critical radius, in the sense that there exists a value $R_c$ above which $p=1$. Considering the mean avalanche size $<s>$ for several $R$, we have found that there is a power law divergence at $R_c$. In other words, for $R$ higher than $R_c$ no jamming can happen. The value of this critical radius is $R_c = 4.94 \pm 0.03$ for spherical beads. This holds true for different materials. This means that there is no influence in the jamming of the material properties of the grains. However, a small change in the grain shape gives rise to a significant change in the critical radius: if the grains are not spherical there is still a critical radius, but its value depends on the grain shape. Interestingly, we have always found approximately the same critical exponent for the
divergence of $<s>$. We therefore conclude that the material properties of individual grains that we have explored do not have any significant influence on arch formation. 
The same cannot be asserted of the friction coefficient because this parameter has not been changed significantly for spherical gains. This issue merits further investigation.

It would be interesting to characterize this jamming to non-jamming transition with an ``order parameter'', in the thermodynamic sense, that is zero for one phase and different from zero in the other. More work is needed to advance in the understanding of how the geometry of the grains influences the shape of the histogram and the jamming probability. The behaviour of small avalanches also merits further investigation.  

\section{Acknowledgements}

J.M. Pastor and I. Zuriguel thank the Asociaci\'on de Amigos de la Universidad de Navarra for a fellowship. 
The Spanish MCYT (Project BFM2002-00414 and HF2002-0015), the Gobierno de Navarra and the Universidad de Navarra (PIUNA project) have supported this work. L.A. Pugnaloni is a member of CONICET (Argentina). 

%We are grateful to Jean-François Boudet, Yacine Amarouchene and Hamid Kellay (CPMOH, University of Bordeaux) for their suggestions and comments.

\newpage
\vspace{-0.5cm}
\begin{small}

\end{small}

\pagebreak

\newpage
\begin{center}
%**********************************************
%\textbf{ Figure Captions}
%\begin{itemize}
%\item[]\textbf{ Figure 1:} 
%\item[]\textbf{ Figure 2:}  
%\item[]\textbf{ Figure 3:}
%\item[]\textbf{ Figure 4:}
%\end{itemize}

\begin{table}
\caption{\label{tab:table1} Properties of the different grains used in this work. $e$ is the restitution coefficient and $\theta$ is the angle at which an avalanche develops in a pile of grains. Note that for non spherical beads $r_s$ and $r_b$ are the small and large radius respectively. $r_{eq}$ is the sphere equivalent radius for the volume of each grain. For spherical beads $r_{eq}$ is the radius of the sphere. The errors are the standard deviations of the results obtained for several measurements.}
%The density of the grains has been obtained by measuring the weight and the real volume of a group of more than $10^5$ grains.
\vspace{1cm}
\begin{ruledtabular}
\begin{tabular}{c|c|c|c|c|c|c|c|c}

\;set\; & material &weight (mg)& $\rho$ $(g/cm^3)$ &     $r_s$ (mm)     &   $r_l$ (mm)   &    $r_{eq}$    & $e$ & $\theta$ $(^o)$ \\[0.5ex]
1 & glass & 10.1 $\pm$ 0.3 & 2.2 $\pm$ 0.1 & & & 1.03 $\pm$ 0.01 & 0.97 $\pm$ 0.03 & 26 $\pm$ 1 \\
2 & glass & 34.7 $\pm$ 0.4 & 2.4 $\pm$ 0.1 & & & 1.52 $\pm$ 0.01 & 0.97 $\pm$ 0.03 & 26 $\pm$ 1 \\
3 & glass & 1.27 $\pm$ 0.4 & 2.4 $\pm$ 0.1 & & & 0.52 $\pm$ 0.005 & 0.97 $\pm$ 0.03 & 27 $\pm$ 1 \\
4 & glass & 11.1 $\pm$ 3.9 & 2.4 $\pm$ 0.1 & & & 1.03 $\pm$ 0.12 & 0.97 $\pm$ 0.03 & 29 $\pm$ 1\\
5 & glass & 26.0 $\pm$ 2.1 & 2.5 $\pm$ 0.1 & & & 1.35 $\pm$ 0.04 & 0.97 $\pm$ 0.03 & 26 $\pm$ 1\\
6 & lead & 46.0 $\pm$ 3.8 & 11.4 $\pm$ 0.5 & & & 0.99 $\pm$ 0.03 & 0.49 $\pm$ 0.10 & 25 $\pm$ 1\\
7 & lead & 150 $\pm$ 14 & 10.9 $\pm$ 0.5 & & & 1.49 $\pm$ 0.05 & 0.49 $\pm$ 0.09 & 27 $\pm$ 1 \\
8 & delrin & 18.9 $\pm$ 0.3 & 1.34 $\pm$ 0.05 & & & 1.50 $\pm$ 0.01 & 0.92 $\pm$ 0.02 & 29 $\pm$ 1\\
9 & steel & 3.98 $\pm$ 0.1 & 7.60 $\pm$ 0.3 & & & 0.50 $\pm$ 0.005 & 0.97 $\pm$ 0.03 & 27 $\pm$ 1\\ 
10& pasta & 5.95 $\pm$ 0.5 & 1.7 $\pm$ 0.2 & 0.92 $\pm$ 0.1 & 0.97 $\pm$ 0.01 & 0.95 $\pm$ 0.1 & & 31 $\pm$ 1 \\ 
11& lentils & 33 $\pm$ 5 & 1.3 $\pm$ 0.5 & 1.22 $\pm$ 0.02 & 2.22 $\pm$ 0.2 & 1.81 $\pm$ 0.12 & & 38 $\pm$ 1 \\
12& rice & 15.9 $\pm$ 2.9 & 1.2 $\pm$ 0.4 & 0.98 $\pm$ 0.11 & 3.3 $\pm$ 0.4 & 1.47 $\pm$ 0.15 & & 42 $\pm$ 1 \\ 
%\footnotesize{Table 1. Properties of the diferent materials used in this work}
\end{tabular}
\end{ruledtabular}
\end{table}

\begin{figure}[th]
\includegraphics{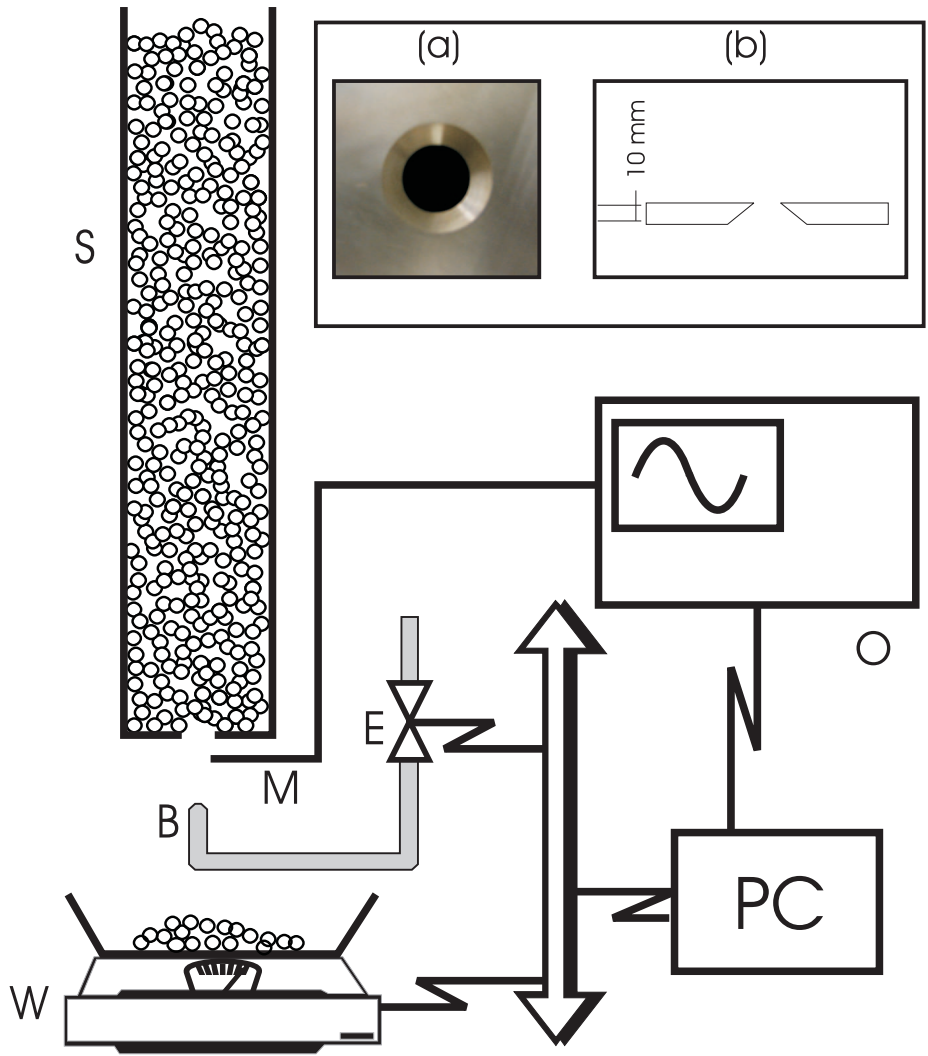}
\caption{\label{sketch} 
Sketch of the experimental set-up. S, silo; W, electronic balance; B, blower; E, electrovalve; M, microphone; O, oscilloscope; PC, computer. In a) a photograph of the orifice is shown as seen from below. In b) a section of the orifice is shown.}
\end{figure}

\begin{figure}[th]
\centering
\begin{minipage}[b]{11cm}
\centering
\includegraphics{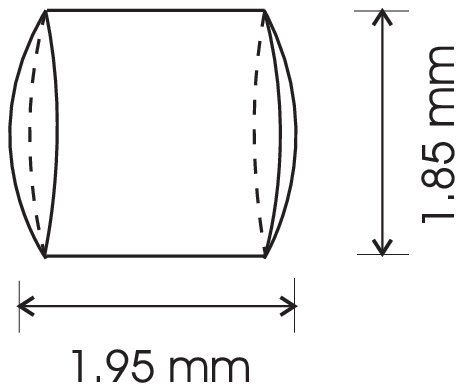}
\end{minipage}
\begin{minipage}[b]{11cm}
\centering
\hspace{1cm}
\includegraphics{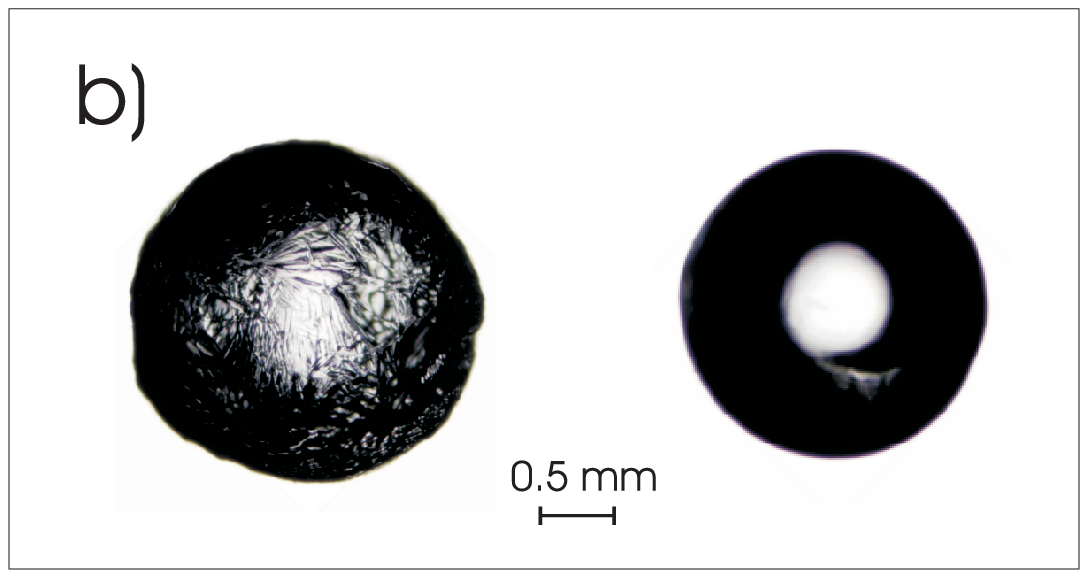}
\end{minipage}
\caption{\label{sis_exp2} 
a) A diagram of the pasta grains and b) two photographs of the glass beads before (rigth) and after (left) the treatment with fluorhidric acid (series 2 and 5 respectively).}
\end{figure}

\begin{figure}[th]
\centering
\begin{minipage}[b]{11cm}
\centering
\includegraphics{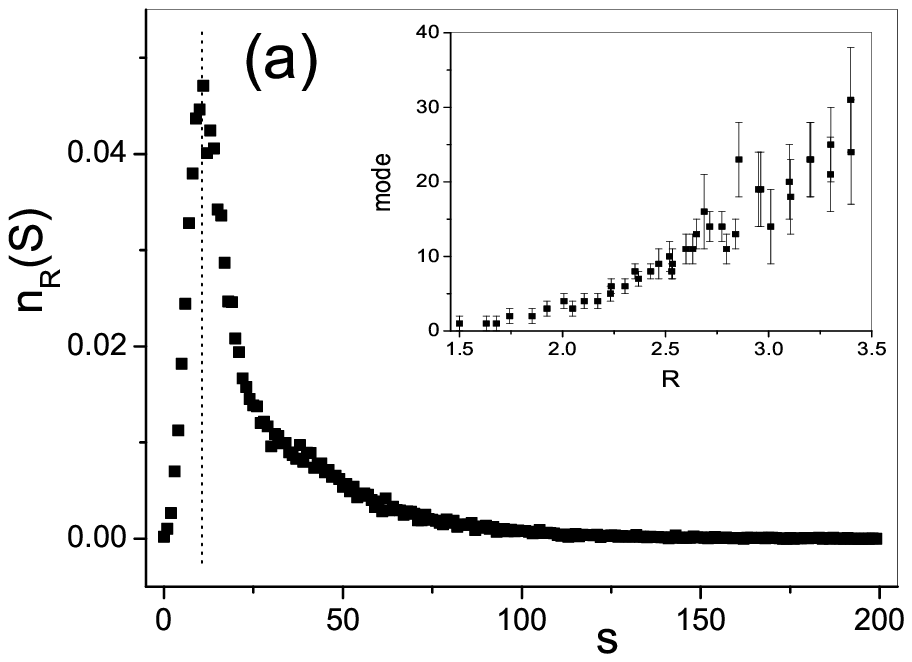}
%\hspace{-0.3cm}
\end{minipage}
\begin{minipage}[b]{11cm}
\centering
\includegraphics{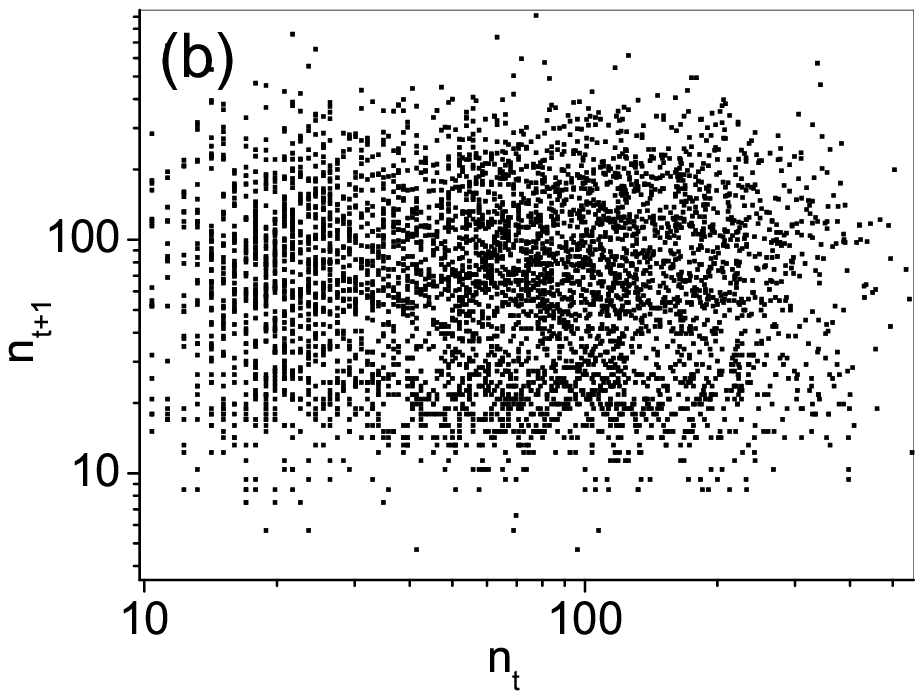}
\end{minipage}
\caption{\label{jam_4histos} 
(a) Histogram for the number of grains $s$ that fall between two successive jams. Data correspond to $R=3$ (beads have a diameter of 2 mm, and the circular orifice is 6 mm wide). The two different regimes are separated with a vertical dashed line. In the inset the position of the mode for different $R$. (b) First return map is plotted for a series of avalanches, \emph{i.e.} the avalanche size $n_t$ versus the next avalanche size $n_t+1$. Note that $t$ is just a correlative index ordering the sequence of avalanches. 
}
\end{figure}

\begin{figure}[th] 
\centering
\begin{minipage}[b]{11cm}
\centering
\includegraphics{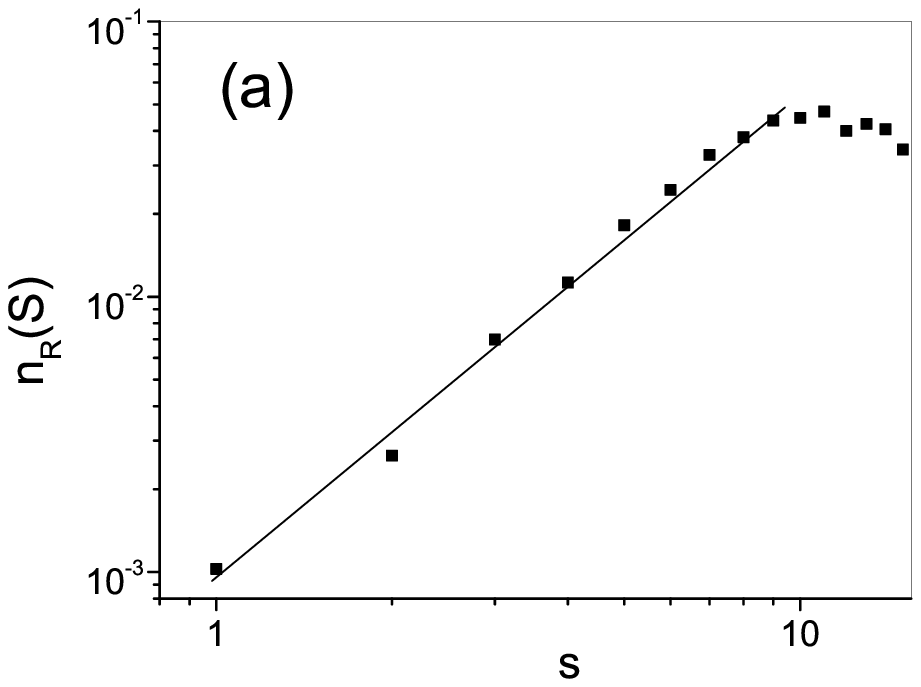}
%\hspace{-0.3cm}
\end{minipage}
\begin{minipage}[b]{11cm}
\centering
\includegraphics{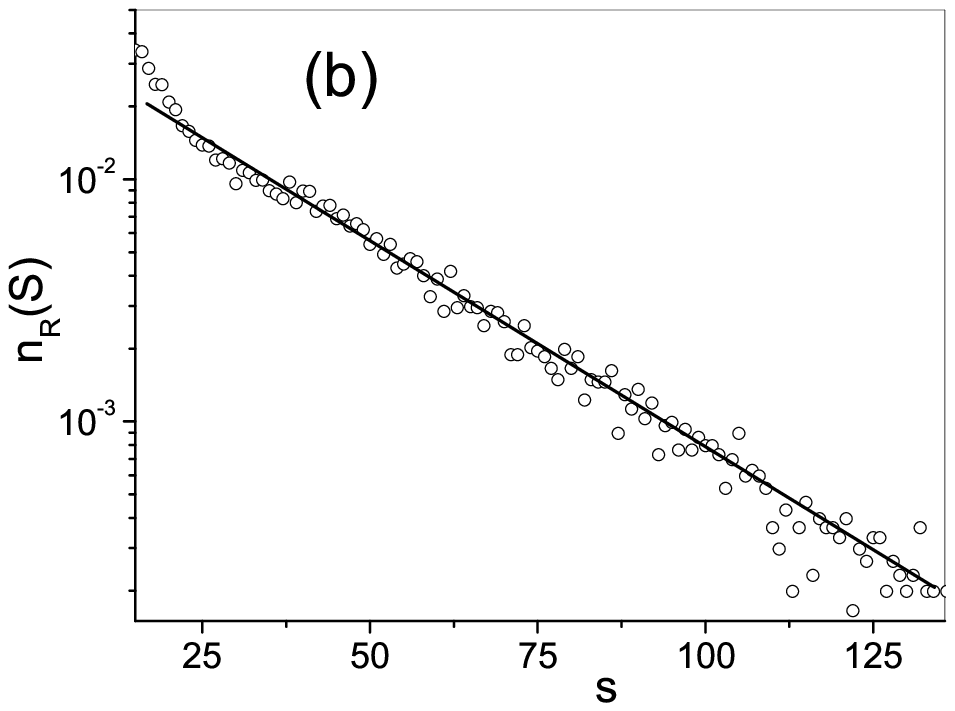}
%\hspace{0.3cm}
\end{minipage}
\caption{\label{jam_4histos2} 
The two regimes marked in Fig.\ref{jam_4histos}a are shown in two separate graphs. In (a) -- in logarithmic scale -- a power law fits the avalanches smaller than the mode. In (b) the exponential tail is shown in semilogarithmic scale.  
}
\end{figure}

\begin{figure}[th] 
\includegraphics{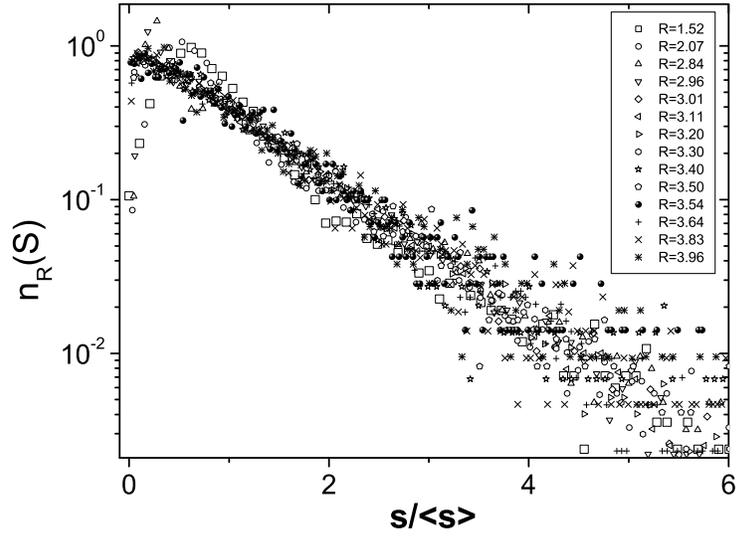}
\caption{\label{histos_nor}
Histogram for the number of grains of the avalanches normalized with the mean avalanche size $<s>$. Different symbols correspond to fourteen histograms from $R=1.52$ to $R=3.96$ that are plotted together. It can be seen that all the histograms collapse into a single curve.}
\end{figure}

\begin{figure}[th] 
\includegraphics{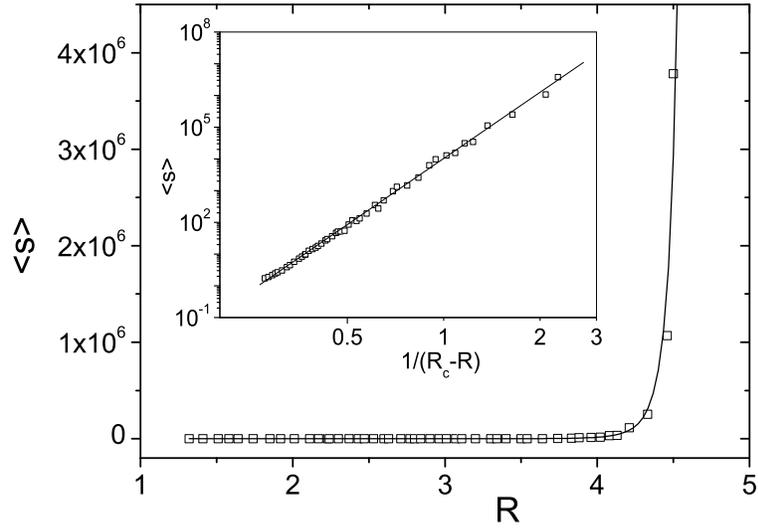}
\caption{\label{s_vs_r}
Mean avalanche size $<s>$ versus $R$. The symbols $\Box$ represent experimental points obtained from the histograms. The solid line is the fit with Eq. \ref{eq3}. Inset: mean avalanche size $<s>$ versus $1/(R_c-R)$. Note the logarithmic scale.}
\end{figure}

\begin{figure}[th] 
\centering
\begin{minipage}[b]{11cm}
\centering
\includegraphics{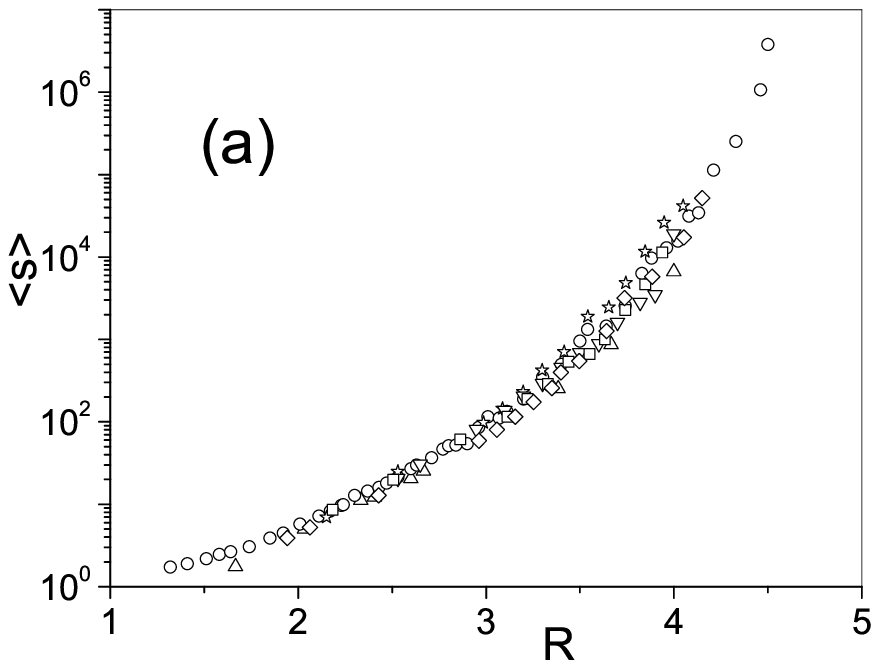}
%\hspace{-0.3cm}
\end{minipage}
\begin{minipage}[b]{11cm}
\centering
\includegraphics{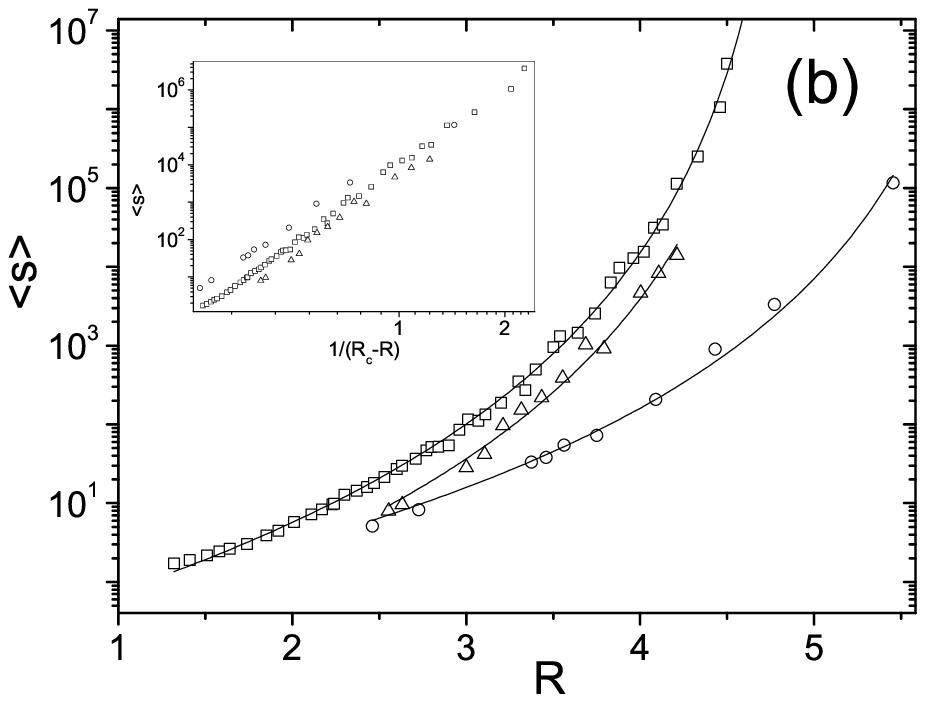}
%\hspace{0.3cm}
\end{minipage}
\caption{\label{diver_mat}
(a) Mean avalanche size $<s>$ versus $R$ in semilogarithmic scale for spherical grains of different types: Delrin ($\bigtriangleup$), glass of two and three mm in diameter ($\bigcirc$), lead of 2 and 3 mm ($\star$), steel ($\bigtriangledown$), rough glass ($+$), and glass with high size dispersion ($\Box$). (b) Mean avalanche size $<s>$ versus $R$ in semilogarithmic scale for different grain shapes. The symbols $\Box$, $\bigcirc$ and $\bigtriangleup$ are experimental points for spheres, rice and pasta grains respectively. The solid line is the fit with Eq. \ref{eq3} with $\gamma=6.909$, which remains constant for all the shapes, while $R_c$ and $A$ change considerably. This can be observed in the inset, where the mean avalanche size $<s>$ versus $1/(R_c-R)$ is plotted in logarithmic scale.}
\end{figure}

\begin{figure}[th] 
\centering
\begin{minipage}[b]{12cm}
\centering
\includegraphics{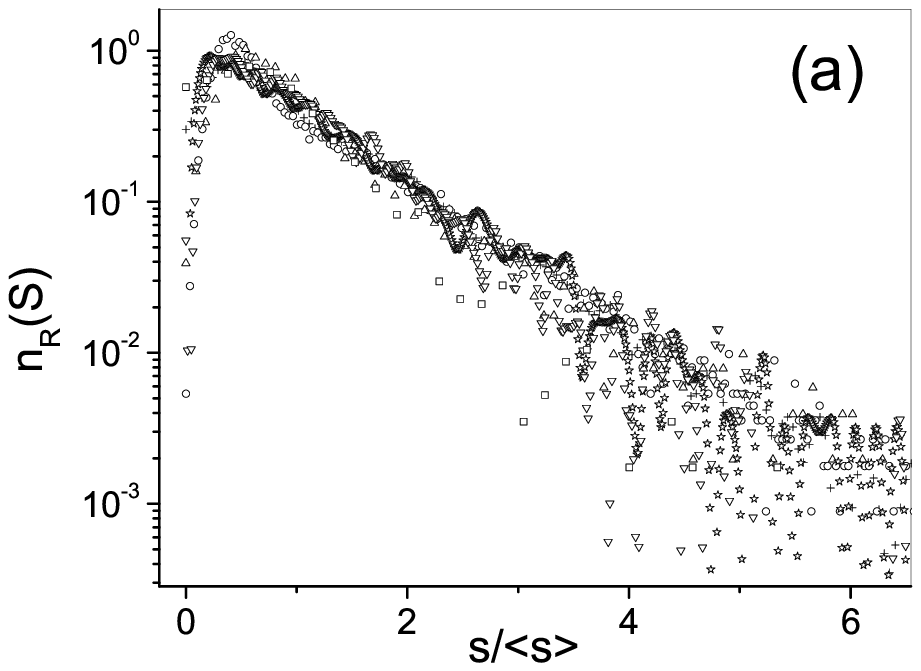}
%\hspace{-0.3cm}
\end{minipage}
\begin{minipage}[b]{12cm}
\centering
\includegraphics{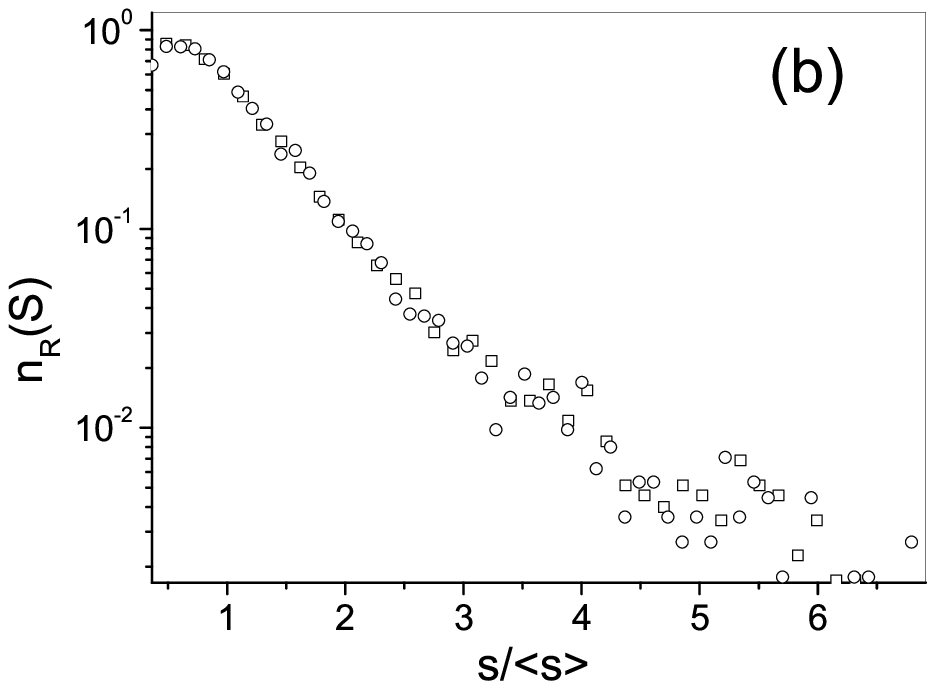}
\end{minipage}
\caption{\label{histos_mat}
(a) Normalized histogram for spherical grains with different material properties. Delrin ($\bigtriangleup$), glass of two and three mm in diameter ($\bigcirc$), lead of 2 and 3 mm ($\star$), steel ($\bigtriangledown$), rough glass ($+$), and glass with high size dispersion ($\Box$). All the histograms were obtained for $R$ between 2 and 3. (b) Normalized histogram for rice ($\Box$) and lentils ($\bigcirc$) with $R=2.72$ and $R=2.2$ respectively. Note that both figures are plotted in semilogarithmic scale.
}
\end{figure}

\begin{figure}[th] 
\includegraphics{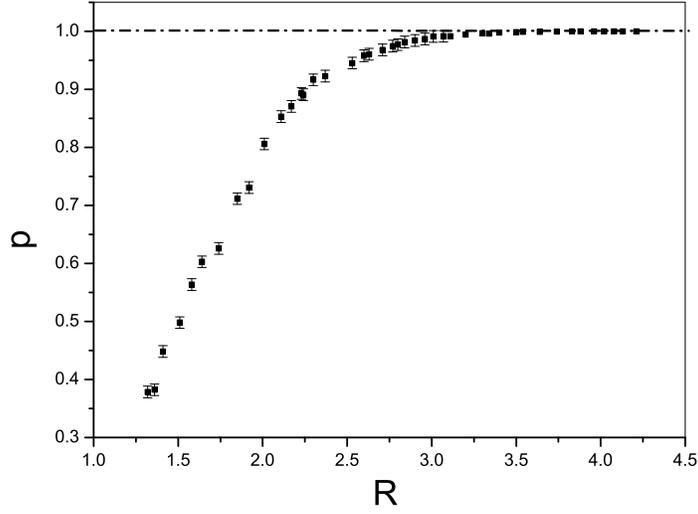}
\caption{\label{figurap}
Probability that one grain past through the outlet ($p$) versus $R$.} 
\end{figure}

\begin{figure}[th] 
\includegraphics{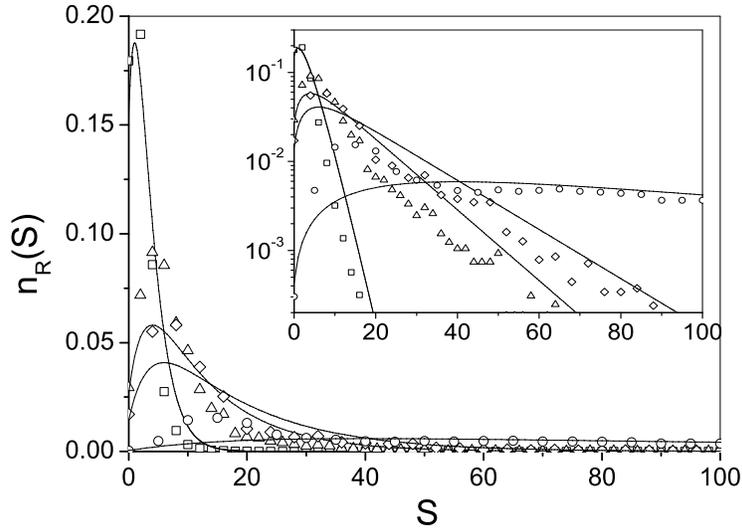}
\caption{\label{figura_luis}
The avalanche size distribution for spherical glass beads. $\Box$, $\bigtriangleup$, $\Diamond$ and $\bigcirc$ correspond to $R=1.74$, $2.23$, $2.42$ and $3.07$, respectively. Solid lines correspond to Eq. \ref{e7} (see text for details of the fitting procedure). The corresponding values of the parameters $p$ and $\alpha$ are: $p=0.615$, $\alpha=1.94$ ($R=1.74$); $p=0.913$, $\alpha=2.69$ ($R=2.23$); $p=0.938$, $\alpha=3.70$ ($R=2.42$); and $p=0.992$, $\alpha=18.52$ ($R=3.07$). Inset: the same graph in semi-logarithmic scale.} 
\end{figure}

%\begin{figure}[th] 
%\centering
%\begin{minipage}[b]{11cm}
%\centering
%\includegraphics{flujos_1.eps}
%\hspace{-0.3cm}
%\end{minipage}
%\begin{minipage}[b]{11cm}
%\centering
%\includegraphics{flujos_2.eps}
%\hspace{0.3cm}
%\end{minipage}
%\caption{\label{flujos}
%a) Number of beads versus avalanche duration for $R=3.05$. Squares are experimental points and the solid line is a linear fit. The slope of this fit is the flow in number of beads per second. b) Number of beads versus avalanche duration for $R=8$. Note that a jam is never found for $R=8$. Circles are experimental points and the solid line is a linear fit.
%}
%\end{figure}

%\begin{figure}[th] 
%\includegraphics{flu_R.eps}
%\caption{\label{flu_R}
%Flow ($W_b$) versus $R$ for different glass beads. Note that axes are in logarithmic scale. It is clear that the exponent $\beta$ increases when $R$ is decreased: $\beta =2.5$ for high $R$ and $\beta =3.5$ for $R$ near the region of jamming.
%}
%\end{figure}

\end{center}
\end{document}